\documentclass{article}

\usepackage{arxiv}
\usepackage{multicol}

\usepackage{microtype}
\usepackage{graphicx}
\usepackage{subfigure}
\usepackage{booktabs} 

\usepackage[utf8]{inputenc} 
\usepackage[T1]{fontenc}    
\usepackage{hyperref}       
\usepackage{url}            
\usepackage{booktabs}       
\usepackage{amsfonts}       
\usepackage{nicefrac}       
\usepackage{microtype}      
\usepackage{graphicx}
\usepackage{natbib}
\usepackage{doi}
\usepackage{wrapfig}
\graphicspath{ {./images/} }
\usepackage[rightcaption]{sidecap}
\usepackage{textcomp}

\title{Social Neuro AI: \\Social Interaction as the "dark matter" of AI}

\author{
	\href{https://orcid.org/0000-0000-0000-0000}{\includegraphics[scale=0.06]{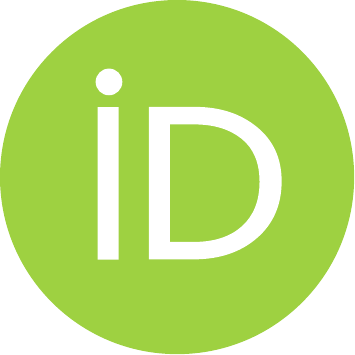}\hspace{1mm}Samuele Bolotta} \\
	Department of Computer Science and Operations Research\\
	Université de Montréal\\
	Montreal, QC, H2S 3H1 \\
	\texttt{samuele.bolotta@ppsp.team} \\
	\And
	\href{https://orcid.org/0000-0000-0000-0000}{\includegraphics[scale=0.06]{orcid.pdf}\hspace{1mm}Guillaume Dumas}\\
	Mila - Quebec Artificial Intelligence Institute\\
	CHU Sainte-Justine Research Center, Department of Psychiatry\\
	Université de Montréal\\
	Montreal, QC, H2S 3H1 \\
	\texttt{guillaume.dumas@ppsp.team} 
}

\hypersetup{
pdftitle={From human social learning to social artificial intelligence},
pdfsubject={Social learning in AI},
pdfauthor={Samuele Bolotta, Guillaume Dumas},
pdfkeywords={Social learning, Reinforcement Learning, Dynamical Systems, Embodied Intelligence},
}

\begin{document}
\maketitle

\vspace{1cm}

\begin{abstract}
This article introduces a three-axis framework indicating how AI can be informed by biological examples of social learning mechanisms. We argue that the complex human cognitive architecture owes a large portion of its expressive power to its ability to engage in social and cultural learning. However, the field of AI has mostly embraced a solipsistic perspective on intelligence. We thus argue that social interactions not only are largely unexplored in this field but also are an essential element of advanced cognitive ability, and therefore constitute metaphorically the “dark matter” of AI. In the first section, we discuss how social learning plays a key role in the development of intelligence. We do so by discussing social and cultural learning theories and empirical findings from social neuroscience. Then, we discuss three lines of research that fall under the umbrella of Social NeuroAI and can contribute to developing socially intelligent embodied agents in complex environments. First, neuroscientific theories of cognitive architecture, such as the global workspace theory and the attention schema theory, can enhance biological plausibility and help us understand how we could bridge individual and social theories of intelligence. Second, intelligence occurs in time as opposed to over time, and this is naturally incorporated by dynamical systems. Third, embodiment has been demonstrated to provide more sophisticated array of communicative signals. To conclude, we discuss the example of active inference, which offers powerful insights for developing agents that possess biological realism, can self-organize in time, and are socially embodied. 
\end{abstract}

\vspace{2cm}

\begin{multicols}{2}

\section{The importance of social learning}

\paragraph{Social learning categories}

Various approaches have been proposed in order to reach a human-like level of intelligence. For example, some argue that scaling foundational models (self-supervised pretrained deep network models), data and compute can lead to such kind of intelligence \cite{yuanzhao, bommasani}. Others argue that attention, understood as a dynamical control of information flow \cite{mittal2020learning}, is all we need. Transformers have proposed a general purpose architecture where inductive biases shaping the flow of information are learned from the data itself \cite{vaswani2017attention}; this architecture can be applied to various domains ranging from sequence learning to visual processing and time-series forecasting. Others argue that by having a complex enough environment, any reward should be enough to elicit some complex behaviour and end up in intelligent behaviour that subserves the maximisation of such reward. This therefore discards the idea that specialised problem formulations are needed for each ability \cite{silver2021}. Our proposal stems from the idea that human cognitive functions such as theory of mind (the capacity to understand other people by ascribing mental states to them) and explicit metacognition (the capacity to reflect on and justify our behaviour to others) are not genetically programmed, but rather constructed during development through social interaction \cite{heyes2}.
. Since their birth, social animals use their conspecifics as vehicles for gathering information that can potentially help them respond efficiently to challenges in the environment, avoiding harm and maximizing rewards \cite{kendal2}. Learning adaptive information from others results in better regulation of task performance, especially by gaining fitness benefits and in avoiding some of the costs associated with asocial, trial-and-error learning, such as time loss and energy loss as well as exposure to predation \cite{clark2016regulation}. Importantly, cultural inheritance permeates a broad array of behavioural domains, including migratory pathways, foraging techniques, nesting sites and mates \cite{whiten2}. The spread of such information across generations gives social learning a unique role in the evolution of culture and therefore makes it a crucial candidate to investigate the biological bases of human cognition \cite{gariepy}. In the current paper, we do not focus extensively on the differences between social learning in humans and in other animals as the cognitive processes used in acquiring behaviour seem to be very similar across a wide range of species \cite{Heyes_2012}. What sets humans apart from other animals, however, is: a) social learning in humans is highly rewarded from early infancy \cite{nielsen} b) the nature of the inputs surrounding humans is way more complex than for other animals \cite{Heyes_2012}. According to the ontogenetic adaptation hypothesis \cite{tomasello}, human infant's unique social-cognitive skills are the result of shared intentionality (capacity to share attention and intention) and are adaptations for life in a cultural group - with individuals coordinating, communicating and learning from each other in several ways. Recent reviews have identified four main categories of social learning that differ in what is socially learnt and in the cognitive skills that are required \cite{hoppitt}, \cite{whiten2} (Figure 1). These categories have been developed through the approach of behaviourism. While we acknowledge that there is more to social learning than mere behaviour (the affective and cognitive dimensions are equally crucial \cite{gruber}), we keep it as the focus of this short article because it is an empirically solid starting point with clarified mechanisms. The purpose of this section, then, is to give an example of social learning mechanisms that are common across multiple species and can be understood as a natural form of Social Neuro-AI. Moreover, this section aims at demonstrating how social interactions are a key component of biological intelligence; we make the case that they might be of inspiration for the development of socially intelligent artificial agents that can cooperate efficiently with humans and with each other. In other words, although there are examples of social agents (chatbots, non-player characters in video games, social robots), we argue that social interactions still remain the "dark matter" of the field. These social behaviours often emerge from a Piagetian perspective on human intelligence. As argued by \cite{kovac}, mainstream Deep Reinforcement Learning research sees intelligence as the product of the individual agent's exploration of the world; it mainly focuses on sensorimotor development and problems involving interaction with inanimate objects rather than social interactions with animate agents. This approach can and has given rise to apparent social behaviours, but we argue that this is not the best approach, as it does not involve any focus on the genuine social mechanisms per se \cite{dumas2014human}. Instead, it sees social behaviours as a collateral effect of the intelligence of a solitary thinker. For this reason, as Schillbach and colleagues (2013) argued a decade ago that social interactions were the "dark matter" of cognitive neuroscience, here we argue that social interactions can also be considered metaphorically as the "dark matter" of AI \cite{schilbach}. Indeed, more than being a rather unexplored topic, social interactions can constitute a critical missing piece for the understanding and modelling of advanced cognitive abilities. 

At the most elementary level, enhancement consists of an agent observing a model that focuses on particular objects or locations and consequently adopting the same focus \cite{heyes1}, \cite{thorpe}. For example, it was demonstrated that bees outside the nest land more often on flowers that they had seen preferred by other bees \cite{worden}. This skill requires social agents to perform basic associative learning in relation to other agents’ observed actions; it is likely to be the most widespread form of social learning across the animal kingdom. A more complex form of social learning consists of observational conditioning, which exposes a social agent to a relationship between stimuli \cite{heyes1}; this exposure causes a change in the agent. For example, the observation of experienced demonstrators facilitated the opening of hickory nuts by red squirrels, relative to trial-and-error learning \cite{weigle}. This is therefore a mechanism through which agents learn the value of a stimulus from the interaction with other agents. Yet a more complex form of social learning consists of affordance learning, which allows a social agent to learn the operating characteristics of objects or environments by observing the behaviour of other agents \cite{whiten2}. For example, pigeons that saw a demonstrator push a sliding screen for food made a higher proportion of pushes than observers in control conditions, thus exhibiting affordance learning \cite{klein}. In other words, the animals perceive the environment partly in terms of the action opportunities that it provides. Finally, at the most complex level, copying another individual can take the shape of pure imitation, where every detail is copied, or emulation, where only a few elements are copied \cite{byrne}. For example, most chimpanzees mastered a new technique for obtaining food when they were under the influence of a trained expert, whereas none did so in a population lacking an expert \cite{whiten}. As to what is required for imitation, there are debates in the literature ranging from the distinctions between program-level and production-level imitation \cite{byrne} to the necessity of pairing Theory of Mind (ToM) with behavioural imitation to obtain ‘true’ imitation \cite{tomasello2005}. We refer the reader to \cite{breazeal} for a more detailed discussion of imitation in robots.

\begin{figure*}[t]
    \centering
    \includegraphics[width=1.0\textwidth]{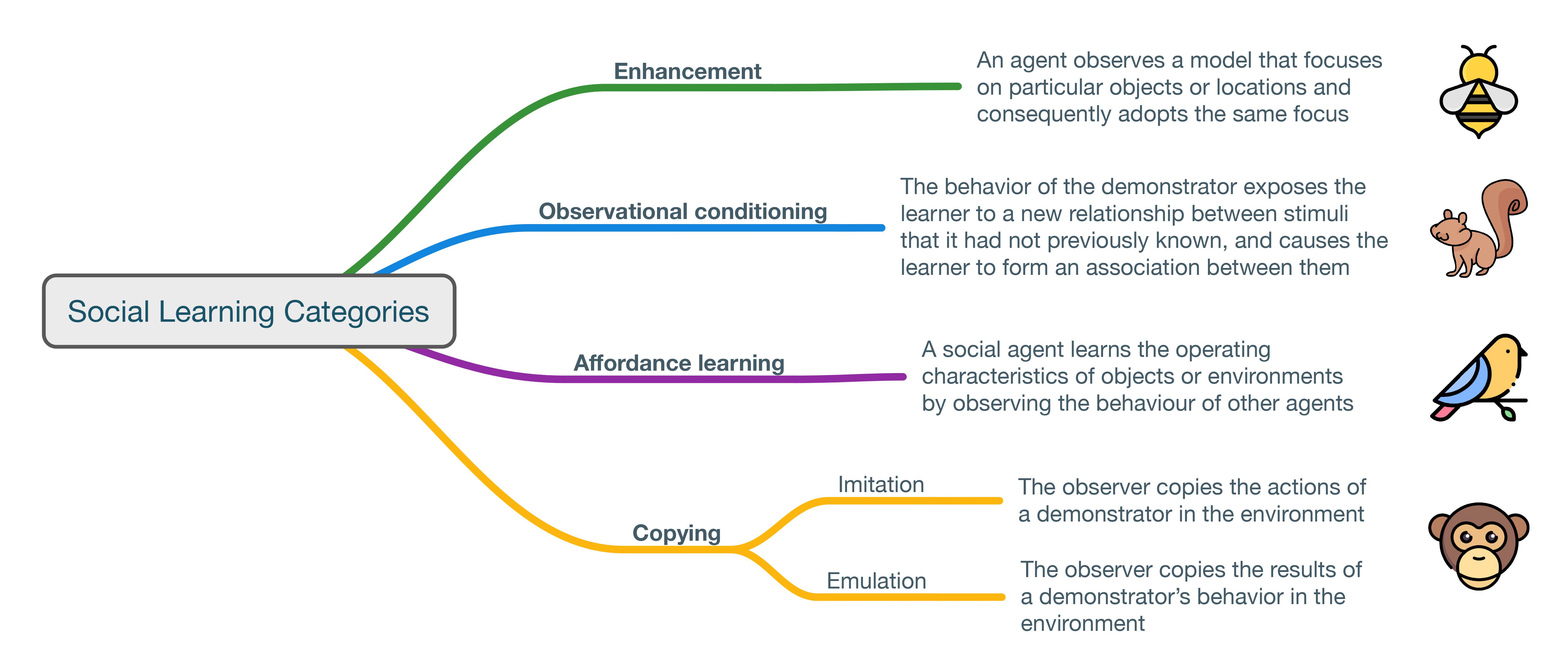}
    \caption{Social learning categories. Figure inspired by \cite{whiten}.}
\end{figure*}

\paragraph{Social learning strategies}
Crucially, while social learning is widespread, using it indiscriminately is rarely beneficial. This suggests that individuals should be selective in what, when, and from whom they learn socially, by following ‘social learning strategies’ (SLSs; \cite{kendal2}). Several SLSs might be used by the same population and even by the same individual. The aforementioned categories of social learning have been shown to be refined by modulating biases that can strengthen their adaptive power \cite{kendal2}. For example, an important SLS is copying when asocial learning would be costly; research has shown that, when task difficulty increases, various animals are more likely to use social information. Individuals also prefer using social information when they are uncertain about a task; high-fidelity copying is observed among children who lack relevant personal information \cite{wood}. In general, other state-based SLSs can affect the decision to use social information, such as age, social rank, and reproductive state of the learner; for example, low- and mid-ranking chimpanzees are more likely to use social information than high-ranking individuals \cite{kendal}. Model-based biases are another crucial category; for example, children prefer to copy prestigious individuals, where status is evidenced by their older age, popularity and social dominance \cite{flynn}. Multiple evidence also suggests that a conformist transmission bias exists, whereby the behaviour of the majority of individuals is more likely to be adopted by others \cite{kendal2}.

\paragraph{Social learning in neuroscience}
We have presented evidence that social learning is a crucial hallmark of many species and it manifests itself across different behavioural domains; without it, animals would lose the possibility to quickly acquire valuable information from their conspecifics and therefore lose fitness benefits. However, one important question is: how does the brain mediate social processes and behaviour? Despite the progress made in social neuroscience and in developmental psychology, only in the last decade, serious efforts have started focusing on the answer to this question - as neural mechanisms of social interaction were seen as the “dark matter” of social neuroscience \cite{schilbach}; recently, a framework for computational social neuroscience has been proposed, in an attempt to naturalize social interaction \cite{tognoli}. At the intra-brain level, it was demonstrated that social interaction is categorically different from social perception and that the brain exhibits different activity patterns depending on the role of the subject and on the context in which the interaction is unfolding \cite{dumas}. At the inter-brain level, functional Magnetic Resonance Imaging (fMRI) or Electroencephalography (EEG) recordings of multiple brains (i.e. hyperscanning) have allowed to demonstrate inter-brain synchronization during social interaction - specifically, while subjects were engaged in spontaneous imitation of hand movements \cite{dumas2}. Interestingly, the increase in coupling strength between brain signals was also shown to be present during a two-person turn-taking verbal exchange with no visual contact, in both a native or a foreign language context \cite{perez}. Inter-brain synchronization is also modulated by the type of task and by the familiarity between subjects \cite{djalovski}. Overall, this shows that, beyond their individual cognition, humans are also coupled in the social dimension. Interestingly, the field of computational social neuroscience has also focused on explaining the functional meaning of such correlations between inter-brain synchronization and behavioural coupling. A biophysical model showed that the similarity of both endogenous dynamics and anatomical structure might facilitate inter-individual synchronization and explain our propensity to socially bind with others via perception and actions \cite{dumas}. More specifically, the connectome, a wiring diagram that maps all neural connections in the brain, not only facilitates the integration of information within brains, but also between brains. In those simulations, tools from dynamical systems thus suggest that beyond their individual cognition, humans are also dynamically coupled in the social realm \cite{dumas}.

\paragraph{Social learning and language development}
Regarding language development in humans, cognitive and structural accounts of language development have often conceptualized linguistic abilities as static and formal sets of knowledge structures, ignoring the contextual nature of language. However, good communication must be tailored to the characteristics of the listener and of the context - language can also be explained as a social construct \cite{whitehurst}. For example, evidence shows that the language outcome of children with cochlear implants is heavily influenced by parental linguistic input during the first years after cochlear implantation \cite{holzinger}. In terms of specific social learning variables, imitation has also been shown to play a major role in boosting language development, usually in the form of selective imitation \cite{whitehurst2, whitehurst3}. Moreover, in children with autism spectrum disorder, social learning variables such as joint attention, immediate imitation, and deferred imitation have been shown to be the best predictors of language ability and rate of communication development \cite{toth}. These results clearly suggest that social learning skills have an influence on language acquisition in humans. 

\section{Steps towards Social Neuro AI}

\paragraph{How could social learning be useful for AI?}

In the previous sections, we have provided convincing evidence that interpersonal intelligence enhances intrapersonal intelligence through the mechanisms and biases of social learning. It is a crucial aspect of biological intelligence that possesses a broad array of modulating biases meant to strengthen its adaptive power. Recent efforts in computational social neuroscience have paved the way for a naturalization of social interactions. 

Multi-agent reinforcement learning (MARL) is the best subfield of AI to investigate the interactions between multiple agents. Such interactions can be of three types: cooperative games (all agents working for the same goal), competitive games (all agents competing against each other), and mixed motive games (a mix of cooperative and competitive interactions). At each timestep t, each agent is attempting to maximize its own reward by learning a policy that optimizes the total expected discounted future reward. We refer the reader to high-quality reviews that have been written on MARL \cite{wong, nguyen, hernandez}. Here, we highlight that, among others, low sample efficiency is one of the greatest challenges for MARL, as millions of interactions with the environment are usually needed for agents to learn. Moreover, multi-agent joint action space increases exponentially with the number of agents, leading to problems that are often intractable. In the last few years, part of the AI community has already started demonstrating that these problems can be alleviated by mechanisms that allow for social learning \cite{jaques, ndousse}. For example, rewarding agents for having a causal influence over other agents' actions leads to enhanced coordination and communication in challenging social dilemma environments \cite{jaques2} and rewarding agents for coordinating attention with another agent improves their ability of coordination, by reducing the cost of exploration \cite{lee2}. More in general, concepts from complex systems such as self-organization, emergent behavior, swarm optimization and cellular systems suggest that collective intelligence could produce more robust and flexible solutions in AI, with higher sample efficiency and higher generalization \cite{hatang}. In the following sections, we argue that to exploit all benefits that social learning can offer AI and robotics, more focus on biological plausibility, social embodiment and temporal dynamics is needed. Studies have focused on the potential of conducting research at the intersection of some of these three axes \cite{husbands, Kerzel_Strahl_Magg_Navarro-Guerrero_Heinrich_Wermter_2017}. Moreover, it is worth noticing that \cite{dumas2012anatomical, heggli2019} offer a tentative glimpse of what the intersection of the three axes would look like - both using dynamical systems with computational simulations to address falsifiable scientific questions associated with the idea of social embodiment.

\subsection{Biological plausibility}
Biological plausibility refers to the extent to which an artificial architecture takes inspiration from empirical results in neuroscience and psychology. The social learning skills and biases that we have shown so far are boosted in humans by their advanced cognitive architecture \cite{whiten2}. Equipping artificial agents with complex social learning abilities will therefore require more complex architectures that can handle a great variety of information efficiently. This is exactly what "Neuro-AI" aims at: drawing on how evolution has shaped the brain of humans and of other animals in order to create more robust agents (Figure 2). While the human unconscious brain aligns well with the current successful applications of deep learning, the conscious brain involves higher-order cognitive abilities that perform much more complex computations than what deep learning can currently do \cite{bengio}. More specifically, “unconsciousness” is where most of our intelligence lies and involves unconscious abilities related to view-invariance, meaning extraction, control, decision-making and learning; “i-consciousness” is the part of human consciousness that is focused on integrating all available evidence to converge toward a single decision; “m-consciousness” is the part of human consciousness that is focused on reflexively representing oneself, utilizing error detection, meta-memory and reality monitoring \cite{graziano}. Notably, recent efforts in the deep learning community have indeed focused on Neuro-AI: building advanced cognitive architectures that are inspired from neuroscience. In particular, the global workspace theory (GWT) is the most widely accepted theory of consciousness, and it postulates that when a piece of information is selected by attention, it may non-linearly achieve “ignition”, enter the global workspace (GLW) and be shared across specialized cortical modules, therefore becoming conscious \cite{baars}; \cite{dehaene}. The use of such a communication channel in the context of deep learning was explored for modelling the structure of complex environments. This architecture was demonstrated to encourage specialization and compositionality and to facilitate the synchronization of otherwise independent specialists \cite{goyal}. Moreover, inductive biases inspired by higher-order cognitive functions in humans have been shown to improve OOD generalization. Overall, this section proposes that we draw inspiration from one structure we know is capable of comprehensive intelligence capable of perception, planning, and decision making: the human brain (Figure 2). For a more extensive discussion on biological plausibility in AI, we refer the reader to \cite{Hassabis_Kumaran_Summerfield_Botvinick_2017, macpherson}.

\subsection{Temporal dynamics} (Figure 2). More specifically, FFNs allow signals to travel only from input to output, whereas RNNs can have signals traveling in both directions and therefore introduce loops in the network. Incorporating differential equations in a RNN (continuous-time recurrent neural network) can help learn long-term dependencies \cite{chang2019} and model more complex phenomena, such as the effects of incoming inputs on a neuron. Moreover, viewing RNNs as a discretization of ordinary differential equations (ODEs) driven by input data has led to gains in reliability and robustness to data perturbations \cite{Lim_Erichson_Hodgkinson_Mahoney_2021}. This becomes clear when one notices that many fundamental laws of physics and chemistry can be formulated as differential equations. In general, differential equations are expected to contribute to shifting the perspective from representation-centered to self-organizing agents \cite{brooks}. The former view has been one predominant way of thinking about autonomous systems that exhibit intelligent behaviour: such autonomous agents use their sensors to extract information about the world they operate in and use it to construct an internal model of the world and therefore rationally perform optimal decision making in pursuit of some goal. In other words, autonomous agents are information processing systems and their environment can be abstracted away as the source of answers to questions raised by the ongoing agents' needs. Cognitive processes are thought to incorporate representational content and to acquire such contents via inferential processes instantiated by the brain. Importantly, according to this view, the sensorimotor connections of the agents to the environment are still relevant to understand their behaviour, but there is no focus on what such connections involve and how they take place \cite{newell}. The latter view, in line with the subsumption architecture introduced by Rodney Brooks \cite{brooks}, shows how the representational approach ignores the nonlinear dynamical aspect of intelligence, that is, the temporal constraints that characterize the interactions between agent and environment. Instead, dynamics is a powerful framework that has been used to describe multiple natural phenomena as an interdependent set of coevolving quantitative variables \cite{vangelder} and a crucial aspect of intelligence is that it occurs in time and not over time. If we abstract away the richness of real time, then we also change the behaviour of the agents \cite{smithers}. In other words, one should indeed focus on the structural complexity and on the algorithmic computation the agents need to carry out, but without abstracting away the dynamical aspects of the agent-environment interactions: such dynamical aspects are pervasive and, therefore, necessary to explain the behaviour of the system \cite{vangelder}, \cite{smithers}, \cite{barandiaran}. 

\subsection{Social embodiment}
There has been a resurgence of enactivism in cognitive neuroscience over the past decade, emphasizing the circular causality induced by the notion that the environment is acting upon the individual and the individual is acting upon the environment. To understand how the brain works, then one has to acknowledge that it is embodied \cite{clark}, \cite{hohwy}. Evidence for this shows that embodied intelligence in human children arises from the interaction of the child with the environment through a sensory body that is capable of recognizing the statistical properties of such interaction \cite{smith}. Moreover, higher primates interpret each other as psychological subjects based on their bodily presence; social embodiment is the idea that the embodiment of a socially interactive agent plays a significant role in social interactions. It refers to “states of the body, such as postures, arm movements, and facial expressions, that arise during social interaction and play central roles in social information processing.” \cite{thompson}, \cite{barsalou}. This includes internal and external structures, sensors, and motors that allow them to interact actively with the world. We argue that robots are more socially embodied than digital avatars for a simple reason: they have a higher potential to use parts of their bodies to communicate and to coordinate with other agents (Figure 2). At a high level, sensorimotor capabilities in the avatar and robots are meant to model their role in biological beings: the agent now has limitations in the ways they can sense, manipulate, and navigate its environments. Importantly, these limitations are closely tied to the agent’s function \cite{deng}. The idea of social embodiment in artificial agents is supported by evidence of improvements in the interactions between embodied agents and humans \cite{zhang2016enhanced}. Studies have shown positive effects of physical embodiment on the feeling of an agent's social presence, the evaluation of the agent, the assessment of public evaluation of the agent, and the evaluation of the interaction with the agent \cite{kose}, \cite{gupta}. In robots, social presence is a key component in the success of social interactions and it can be defined as the combination of seven abilities that enhance a robot’s social skills: 1. Express emotion 2. Communicate with high-level dialogue 3. Learn/recognize models of other agents 4. Establish/maintain social relationships 5. Use natural cues 6. Exhibit distinctive personality and character 7. Learn/develop social competencies \cite{lee}. Social embodiment thus equips artificial agents with a more articulated and richer repertoire of expressions, ameliorating the interactions with it \cite{jaques}. For instance, in human-robot interaction, a gripper is not limited to its role in the manipulation of objects. Rather, it opens a broad array of movements that can enhance the communicative skills of the robot and, consequently, the quality of its possible interactions \cite{deng}. The embodied agent is therefore the best model of the aspects of the world relevant to its surviving and thriving, through performing situationally appropriate actions \cite{ramstead} (Figure 2). Therefore, it will be crucial to scale up the realism of what the agents perceive in their social context, going from simple environments like GridWorld to more complex ones powered by video-game engines and, finally, to extremely realistic environments, like the one offered by the MetaHuman Creator of Unreal Engine. In parallel, greater focus is needed on the mental processes supporting our interactions with social machines, so as to develop a more nuanced understanding of what is ‘social’ about social cognition \cite{Cross_Ramsey_2021} and to gather insights critical for optimising social encounters between humans and robots \cite{Henschel_Hortensius_Cross_2020}. For a more extensive discussion on embodied intelligence, we refer the reader to \cite{Roy_Posner}. These advancements will hopefully result in more socially intelligent agents and therefore in more fruitful interactions between humans and virtual agents.

\section{Active inference}

The active inference framework represents a biologically realistic way of moving away from rule-governed manipulation of internal representations to action-oriented and situationally appropriate cognition \cite{friston}. More specifically, active inference can be seen as a self-organising process of action policy selection \cite{ramstead}, which a) concerns the selective sampling of the world by an embodied agent, and b) instantiates in a generative model the goal of minimizing their surprise through perception and action \cite{ramstead}. In other words, generative models do not encode exploitable and symbolic structural information about the world, because cognition does not perform manipulation of internal representations, but rather instantiates control systems that are expressed in embodied activity and utilize information encoded in the approximate posterior belief \cite{ramstead}. Interestingly, by grounding GWT within the embodied perspective of the active inference framework, the Integrated World Modeling Theory (IWMT) suggests that conscious experience can only result from autonomous embodied agents with global workspaces that generate integrative models of the world with spatial, temporal and causal coherence \cite{Safron_2020}.

Active inference models are still very discrete in their architectures, especially regarding high-level cognitive aspects, but they may be a good class of models to raise the tension between computation and implementation (Figure 2). Therefore, they only have been able to handle small policies and state–spaces, while also requiring the environmental dynamics to be well known. However,  using deep neural networks to approximate key densities, the agent can scale to more complex tasks and obtain performance comparable to common reinforcement learning baselines \cite{milidge}. Moreover, one advantage of active inference is that the associated biologically inspired architectures predict future trajectories of the agent N steps forward in time, rather than just at the next step. By sampling from these trajectories, the variance of the decision is reduced \cite{milidge}.

Interestingly, by grounding GWT within the embodied perspective of the active inference framework, the Integrated World Modeling Theory (IWMT) suggests that complexes of integrated information and global workspaces can entail conscious experiences if (and only if) they are capable of generating integrative world models with spatial, temporal, and causal coherence. These ways of categorizing experience are increasingly recognized as constituting essential “core knowledge” at the foundation of cognitive development (Spelke and Kinzler, 2007). In addition to space, time, and cause, IWMT adds embodied autonomous selfhood as a precondition for integrated world modeling. 

\section{A detailed proposal: how can increased biological plausibility enhance social affordance learning in artificial agents?}

Attention has become a common ingredient in deep learning architectures. It can be understood as a dynamical control of information flow \cite{mittal2020learning}. In the last decade, transformers have demonstrated how \emph{attention may be all we need}, obtaining excellent performances in sequence learning \cite{vaswani2017attention}, visual processing \cite{dosovitskiy2020image} and time-series forecasting \cite{lim2021temporal}. While transformers proposed a general purpose architecture where inductive biases shaping the flow of information are learned from the data itself, we can imagine a higher-order informational filter built on top of attention: an \emph{Attention Schema (AS)}, namely a descriptive and predictive model of attention. In this regard, the attention schema theory (AST) is a neuroscientific theory that postulates that the human brain, and possibly the brain of other animals, does construct a model of attention: an attention schema \cite{graziano2015}. Specifically, the proposal is that the brain constructs not only a model of the physical body but also a coherent, rich, and descriptive model of attention. The body schema contains layers of valuable information that help control and predict stable and dynamic properties of the body; in a similar fashion, the attention schema helps control and predict attention. One cannot understand how the brain controls the body without understanding the body schema, and in a similar way one cannot understand how the brain controls its limited resources without understanding the attention schema \cite{graziano}. The key reason a higher-order filter on top of attention seems a promising idea for deep learning comes from control engineering: a good controller contains a model of the item being controlled \cite{conant1970}. More specifically, a descriptive and predictive model of attention could help the dynamical control of attention and therefore maximize the efficiency with which resources are strategically devoted to different elements of an ever-changing environment \cite{graziano2017}. Indeed, the performance of an artificial agent in solving a simple sensorimotor task is greatly enhanced by an attention schema, but its performance is greatly reduced when the schema is not available \cite{wilterson}. Therefore, the study of consciousness in artificial intelligence is not a mere pursuit of metaphysical mystery; from an engineering perspective, without understanding subjective awareness, it might not be possible to build artificial agents that intelligently control and deploy their limited processing resources. It has also been argued that, without an attention schema, it might be impossible to build artificial agents that are socially intelligent. This idea stems from the evidence that points at an overlap of social cognition functions with awareness and attention functions in the right temporo-parietal junction of the human brain \cite{mitchell}. It was then proposed that an attention schema might also be used for social cognition, giving rise to an overlap between modelling one’s own attention and modelling others’ attention. In other words, when we attribute to other people an awareness of their surroundings, we are constructing a simplified model of their attention - a schema of others’ attention \cite{graziano2}. Indeed, such a model would enhance the ability of the agent to predict social affordances in real time, which is a goal the field has been trying to achieve in different ways \cite{ardon, shu}. Without a model of others' attention, even if we had detailed information about them, we could not predict their behaviour on a moment-by-moment basis. However, with a component that tracks how and where other agents are focusing their resources in the environment, the probabilities for many affordances in the environment become computable in real time \cite{Graziano_2019}. Specifically, there are three predictions that are investigated in this proposal. The first prediction is that, without an attention schema, attention is still possible, but it suffers deficits in control and thus leads to worse performance. The second prediction is that an attention schema is useful for modeling the attention of other agents as well \textemdash as the machinery that computes information about other people’s attention is the same machinery that computes information about our own attention \cite{graziano2011}. The third prediction is that an agent equipped with an attention schema is going to have better OOD generalization than a classic Proximal Policy Optimization agent \cite{Schulman_Wolski_Dhariwal_Radford_Klimov_2017}, especially in environments in which the ability to intelligently control and deploy limited processing resources is necessary.

\section{Conclusion}

At the crossroads of robotics, computer science, psychology, and neuroscience, one of the main challenges for humans is to build autonomous agents capable of participating in cooperative social interactions. This is important not only because AI will play a crucial role in daily life well into the future, but also because, as demonstrated by results in social neuroscience and evolutionary psychology, intrapersonal intelligence is tightly connected with interpersonal intelligence, especially in humans \cite{dumas2014your}. In this opinion article, we have proposed an approach that unifies three lines of research that, at the moment, are separated from each other; in particular, we have proposed three research directions that are expected to enhance efficient exchange of information between agents. Biological plausibility attempts to increase the robustness and OOD generalization of algorithms by drawing on knowledge about biological brains; temporal dynamics attempts to better capture long-term temporal dependencies; social embodiment proposes that states of the body that arise during social interaction play central roles in social information processing. Unifying these axes of research would contribute to creating agents that are able to cooperate efficiently in extremely complex and realistic environments \cite{dennis}, while interacting with other embodied agents and with humans. 

\end{multicols}

\begin{figure}[h]
    \centering
    \includegraphics[width=1.0\textwidth]{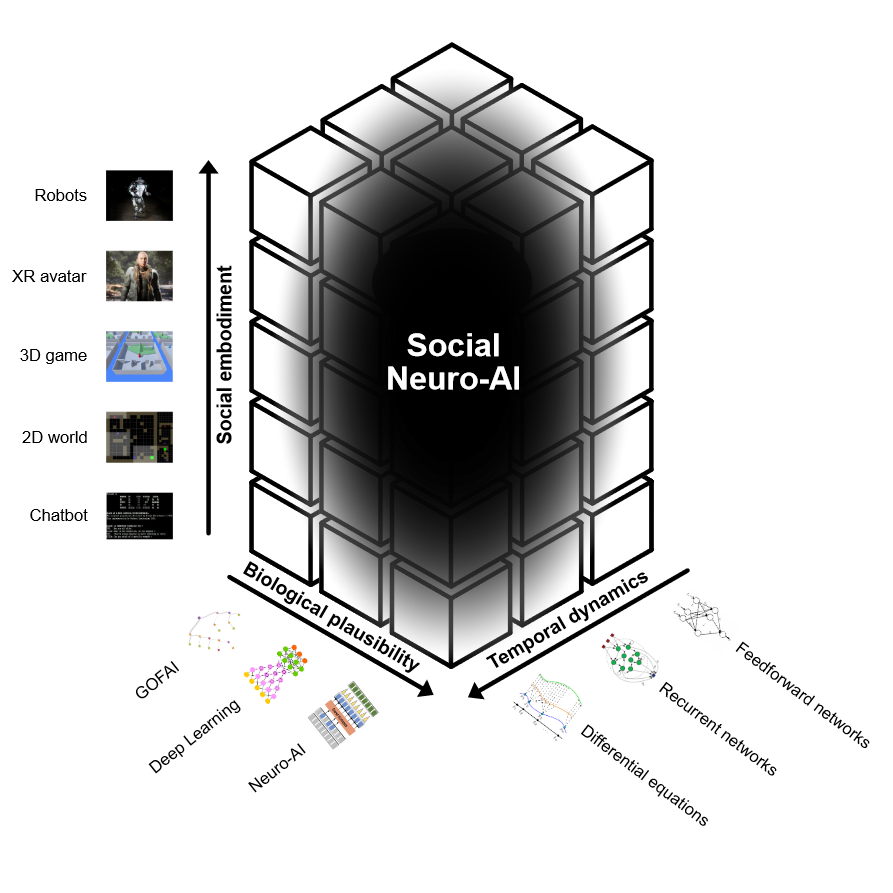}
    \caption{Billions of humans interact daily with algorithms — yet AI is far from human social cognition. We argue that creating such socially aware agents may require "Social Neuro-AI" — a program developing 3 research axes: 1. Biological plausibility 2. Temporal dynamics 3. Social embodiment. Overall, those steps towards socially aware agents will ultimately help in aligned interactions between natural and artificial intelligence. Figure inspired by \cite{schilbach}.}
\end{figure}

\clearpage
\begin{multicols}{2}
\bibliography{references.bib} 
\end{multicols}

\end{document}